\documentclass[graybox]{svmult}

\usepackage{type1cm}        
\usepackage{makeidx}         
\usepackage{graphicx}        
\usepackage{multicol}        
\usepackage[bottom]{footmisc}

\usepackage{newtxtext}       %
\usepackage[varvw]{newtxmath}       
\usepackage{physics}
\usepackage{quantikz}
\usepackage{tikz}
\usepackage{amsmath}

\makeindex             

\begin{document}

\title*{Efficient State Preparation for Quantum Machine Learning}
\author{Chris Nakhl, Maxwell West, Muhammad Usman}
\institute{Chris Nakhl \at School of Physics, The University of Melbourne, Parkville, Victoria 3010, Australia \\
\email{chris.nakhl@unimelb.edu.au} \\
Maxwell West \at School of Physics, The University of Melbourne, Parkville, Victoria 3010, Australia \\
Muhammad Usman \at School of Physics, The University of Melbourne, Parkville, Victoria 3010, Australia \at Data61, CSIRO, Clayton, Victoria 3168, Australia 
}
%
%
\maketitle

\abstract*{One of the key considerations in the development of Quantum Machine Learning (QML) protocols is the encoding of classical data onto a quantum device. In this chapter we introduce the Matrix Product State representation of quantum systems and show how it may be used to construct circuits which encode a desired state. Putting this in the context of QML we show how this process may be modified to give a low depth approximate encoding and crucially that this encoding does not hinder classification accuracy and is indeed exhibits an increased robustness against classical adversarial attacks. This is illustrated by demonstrations of adversarially robust variational quantum classifiers for the MNIST and FMNIST dataset, as well as a small-scale experimental demonstration on a superconducting quantum device.}

\abstract{One of the key considerations in the development of Quantum Machine Learning (QML) protocols is the encoding of classical data onto a quantum device. In this chapter we introduce the Matrix Product State representation of quantum systems and show how it may be used to construct circuits which encode a desired state. Putting this in the context of QML we show how this process may be modified to give a low depth approximate encoding and crucially that this encoding does not hinder classification accuracy and is indeed exhibits an increased robustness against classical adversarial attacks. This is illustrated by demonstrations of adversarially robust variational quantum classifiers for the MNIST and FMNIST dataset, as well as a small-scale experimental demonstration on a superconducting quantum device.}

\section{Introduction}
\label{sec:1}
Crucial to the implementation of Quantum Machine Learning algorithms on quantum devices is the loading of classical data onto a quantum processor. To do so a number of different encodings exist, such as basis encoding for the encoding of binary data, or angle encoding in which each value in a  vector is assigned a qubit which is rotated proportionally to its value. However, a more intuitive encoding exists in \textit{amplitude encoding} wherein some input vector $v_i$ is encoded onto the quantum processor as $\sum_i v_i\ket{i}$. The inherent advantage of this is that it requires exponentially fewer qubits than those which require one qubit for each feature. The primary set back however is that the quantum circuit to prepare such an arbitrary state is in general costly with a depth scaling as $\mathcal{O}(4^n)$~\cite{shende}, where $n$ is the number of qubits. Wishing to decrease the depth of the quantum circuit without necessitating an increase in the number of qubits, one may turn to heuristic methods such as variational encoding~\cite{drastic} or genetic algorithms~\cite{gasp} which find some approximation to the desired state. Alternatively, a more direct method of constructing these states with possibly low depth may be found by considering the Matrix Product State (MPS) representation of the state which can iteratively find circuits that better approximations the state without relying on heuristics. 

In the following chapter, we will focus on the MPS assisted state preparation method by introducing MPS, their construction given an input vector $v_i$ and the algorithm used construct a circuit that prepares $\ket{v}$ from the MPS representation. We will then briefly summarise the paradigmatic QML algorithm, the Quantum Variational Classifier (QVC) and provide a demonstration of how QVCs are resilient towards noise in the input state and how one can use this to their advantage when preparing circuits for the input state. Finally, we show the intriguing result whereby QVCs trained on noisy input states appear to exhibit an increased resilience against classical adversarial attacks.

\section{Introduction to Matrix Product States}
MPS are a one-dimensional tensor network representation of many-body quantum systems that have found significant application in the simulation of quantum computing~\cite{dang,nakhl,nakhl2,moremps,mpsqnn}, and many-body physical systems~\cite{entmbs,cirac04,mpstomo}. States represented as a MPS have the form 
\begin{equation}
    \ket{\psi} = \sum_{i_1,i_2,\dots,i_N} A_{i_1}^{(1)}A_{i_2}^{(2)}\dots A_{i_N}^{(N)} \ket{i_1i_2\dots i_N}
\end{equation}
where $N$ is the total number of qubits, $i_s\in\{0,1\}$ and $A_{i_s}^{(s)}$ are matrices of size $\chi \times \chi$ where $\chi$ is the bond-dimension of the state\footnote{Note that the matrices at the end of the matrix product are of size $1\times\chi$ and $\chi\times 1$ respectively}. The matrices in the MPS in some sense encode the entanglement of the state, which can be more directly probed by considering the singular values of the matrix. 

MPS, and tensor networks in general have a convenient graphical notation which one would be remiss not to bring up. The building blocks of \emph{Penrose Notation} are solid shapes denoting tensors, with lines protruding from them indicating each of the indices of the tensor. For example a matrix is a tensor with two indices and hence may be represented using a solid shape with two protruding lines. 
\begin{equation*}
   A = a_{ij} = \begin{array}{c}\includegraphics[scale=0.5]{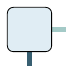}\end{array}
\end{equation*}
Connecting the lines of two tensors in this notation represents summing over an index. For example in matrix multiplication 
\begin{equation*}
   AB = \sum_{j} a_{ij}b_{jk} = a_{ij}b_{jk} = \begin{array}{c}\includegraphics[scale=0.5]{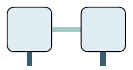}\end{array}
\end{equation*}
For those familiar with the\emph{Einstein summation convention} this representation should look pretty familiar. In this graphical notation then, an MPS may be represented as
\begin{equation}
    \ket{\psi} = \begin{array}{c}
         \includegraphics[scale=0.7]{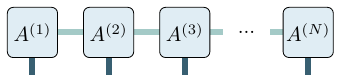}
    \end{array} \label{eq:mps}
\end{equation}
where each tensor $A^{(s)} = \sum_{i=0,1} \ket{i} \otimes A_{i}^{(s)}$

\subsection{Creating the MPS}
Given that we are ostensibly starting off with classical data which we wish to encode, we must set up the protocol which takes some classical input vector $v_i$ and forms an MPS which encodes $\sum_i v_i \ket{i}$. Luckily for us, the tools to do this are already commonplace and hopefully familiar to those who have encountered linear algebra before. The two core operations needed to construct an MPS from a state-vector are matrix decomposition and reshaping. 

There are a number of ways to decompose some matrix into a product of different matrices, but the one that is perhaps most of interest to us is the Singular Value Decomposition (SVD). This decomposition takes some matrix $A$ and returns $A=USV^\dagger$ where $U$ and $V$ are unitary matrices, and $S$ is a diagonal matrix of singular values. Of importance is that if some singular values are sufficiently small then they may be discarded and hence one may be able to reduce the number of columns in $U$ and $V$. This results in a more memory-efficient representation of the state, and will have positive consequences when it comes to determining a circuit representation of the state. 

In order to be able to utilise SVD (or indeed any matrix decomposition) one must of course have a matrix to begin with, this is where reshaping comes in. Fundamentally this is a rather simple operation where one redistributes data over a different set of indices but such that the tensor is still of the same size overall. Computers will use either a row-major or column-major order when reshaping data along a given set of indices, but fundamentally this ordering does not matter to us just that one remains consistent when reading, writing and reshaping data. To start building our MPS we must first reshape the input vector of size $2^N$ to a matrix of size $2\times 2^{N-1}$. Following this one may then use SVD to result in three matrices of size $2 \times s_0$, $s_0\times s_0$ and $s_0\times 2^{N-1}$. The $2\times s_0$ matrix is $U=A^{(0)}$ as per Equation~\eqref{eq:mps}.

To build the rest of the MPS we reshape the $V^\dagger$ matrix to $2s_{i-1} \times 2^{N-i}$ where $i$ counts how many SVDs have been performed thus far and then perform another SVD. The resulting $U$ of size $2s_{i-1} \times s_i$ may be reshaped into $2\times s_{i-1} \times s_i$ and stored as $A^{(i)}$. The $s_i$ are the number of singular values resulting from each SVD and in this context are referred to as the bond-dimension at site $i$. Note that the matrices which store the singular values that result from the SVD may either be absorbed into the $A^{(i)}$ or $A^{(i+1)}$, or simply just left as is and inserted along the internal bond connecting $A^{(i)}$ and $A^{(i+1)}$.

With the MPS now in hand, we may now begin to answer the question of how to we build a low-depth circuit that encodes $v_i$.

\section{State Preparation using MPS} \label{sec:state_prep}

Now that we've gone to all that trouble to build the MPS, we must now destroy it. This is done by determining a series of unitary operators that sequentially disentangle each qubit from the rest of the system. Strictly speaking, your state does not need to be represented as an MPS to perform this procedure, but it does make the implementation more efficient. 

The procedure begins by finding the reduced density matrix $\rho_{0\dots k}$ of the state over the first $k$ sites. For an MPS in canonical form\footnote{An MPS is in a \textit{canonical form}~\cite{orus} when it is left orthogonal at all sites to the left of some site $i$, and right orthogonal to the right of $i$. Site $i$ is refereed to as the the orthogonality centre}, finding the reduced density matrix over some linear bipartition is computationally efficient. We can then find the eigendecomposition of the reduced density matrix as
\begin{equation}
   \rho_{0\dots k} = \sum_{i=0}^{2^k} \lambda_i \ket{i}\bra{i} 
\end{equation}
If $k=\lceil \log(\chi) \rceil + 1$ where $\chi$ is the largest bond dimension of the MPS. We note that for this selection of $k$, the rank of the reduced density matrix is $2^{k-1}$ and therefore there are only $2^{k-1}$ non-zero eigenvalues. Given this decomposition one can construct the unitary operator
\begin{equation}
    U_0 = \sum_{i=0}^{2^k} \ket{i}\bra{\phi_i}
\end{equation}
where the eigenstates are ordered in order of decreasing eigenvalue. When applied to the reduced density matrix $\rho_{0\dots k}$ this gives,
\begin{equation}
    U_o^\dagger\rho U_0= \sum_{i=0}^{2^{k-1}} \lambda_i\ket{i}\bra{i}. \label{eq:disent}
\end{equation}
It is apparent that this operation completely disentangles the first qubit from the rest of the system. This process can then be repeated for the subsystems $1\dots k+1, 2\dots k+2,\dots,N-k\dots N$.  The $U_i$ can then be used to reconstruct the original state,
\begin{equation}
    \ket{\psi} = \sum_i v_i\ket{i}=\prod_{i=N-k+1}^{0}U^\dagger_i \ket{0}.
\end{equation}

This process works well for MPS with low maximum bond dimension, but it will not produce a useful set of unitaries for a highly entangled MPS. The way around this is to select a smaller value of $k$. In this case Equation~\eqref{eq:disent} does not hold, however, provided that the eigenvalues $\lambda_i$ are not all equal at any given step $j$ then it is evident that the probability of the probability of measuring $\ket{0}$ at site $j$ will increase. Hence, at the end of the procedure, we have not completely disentangled the state $\rho$ but have increased the probability of measuring $\ket{0}^{\otimes N}$. One can then repeat the process from the beginning with the new $\rho$ to further destroy the state. After many repetitions one should finally arrive at $\rho = (\ket{0}\bra{0})^{\otimes N}$ and a series of unitaries $U_i^j$ which can be used to reconstruct the original $\rho$. In general, for any small fixed $k$ the number of repetitions required to achieve this perfect disentangling scales as $\mathcal{O}(2^N)$, however one is also free to terminate the algorithm before this point and have a set of unitaries which only approximately prepare the state. To keep track of the fidelity that would be achieved using the unitaries from $j$ iterations, one can keep track of $\bra{0}\rho_j\ket{0}$ where $\rho_j$ here is the state after $j$ iterations of the state preparation algorithm.

\begin{figure}
    \centering
    \begin{tikzpicture}
    \node at (-3.25,3.5) {(a)};
    \node at (0,0) {
    \begin{quantikz}[row sep={2mm}, column sep=0.1cm]
    & \gate[2, style={fill=green!20, rounded corners}]{U_0^0} & &&&&&& & \ldots \\
    & & \gate[2, style={fill=green!20, rounded corners}]{U_1^0} & & &&&& &\ldots\\
    & & &\gate[2, style={fill=green!20, rounded corners}]{U_2^0} & & & && &\ldots\\
    & & &&\gate[2, style={fill=green!20, rounded corners}]{U_3^0} & && && \ldots\\
    & & &&&\gate[2, style={fill=green!20, rounded corners}]{U_4^0} & &&& \ldots\\
    & & &&&&\gate[2, style={fill=green!20, rounded corners}]{U_5^0} & && \ldots\\
    & & &&&&&\gate[2, style={fill=green!20, rounded corners}]{U_6^0} & & \ldots\\
    & & &&&& & & &  \ldots\\
    \end{quantikz}
    };
    \node at (3.5,3.5) {(b)};
    \node at (5.5,0) {
    \begin{quantikz}[row sep={2mm}, column sep=0.1cm]
    & \gate[2, style={fill=red!20, rounded corners}]{U_0^0} & &&&&&& \\
    & & \gate[3, style={fill=red!20, rounded corners}]{U_4^0} &&&& & & & \ldots\\
    & \gate[2, style={fill=red!20, rounded corners}]{U_1^0} & &  &&&&& \\
    & & & \gate[2, style={fill=red!20, rounded corners}]{U_6^0}&&& & & &  \ldots\\
    & \gate[2, style={fill=red!20, rounded corners}]{U_2^0} & \gate[3, style={fill=red!20, rounded corners}]{U_5^0} &&&&&& \\
    & & &&&& & & &  \ldots\\
        & \gate[2, style={fill=red!20, rounded corners}]{U_3^0} & &&&&&& \\
    & & &&&& & & &  \ldots\\
    \end{quantikz}
    };
    \end{tikzpicture}
    \caption{The state preparation circuits resulting from the procedure outlined in Section~\ref{sec:state_prep}. (a) uses the conventional procedure as described with $k=2$ where as (b) is seeking to minimise the depth of the each individual layer. Note that in (b) the ``disentangled'' qubit is the top qubit for unitaries acting on the first four qubit and the bottom qubit elsewhere and that $U_4^0$ and $U_5^0$ are two qubit unitaries acting on the top and bottom qubits only.}
    \label{fig:mps_circ}
\end{figure}
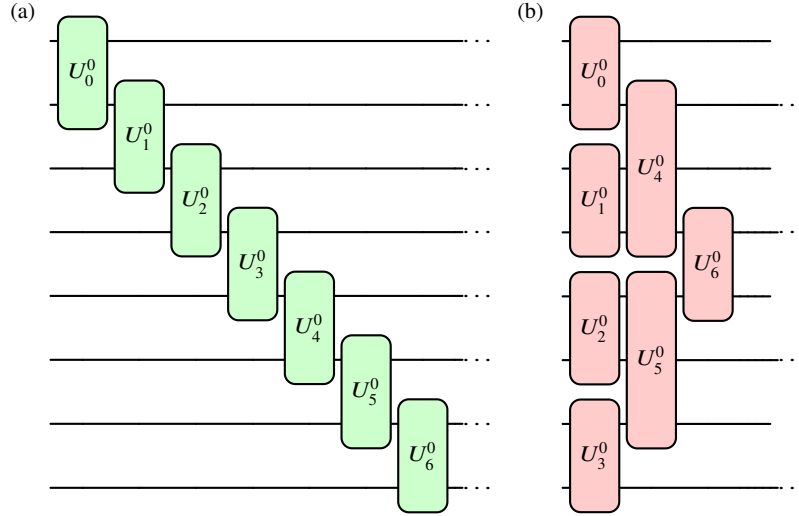

Typically, one will select $k=2$ as the resulting two qubit unitary gates can be decomposed into no more than three CNOT gates and six single qubit rotation gates. The resulting circuit as per Figure~\ref{fig:mps_circ}(a) features only nearest neighbour operations which may be somewhat beneficial depending on the quantum hardware architecture available. Indeed, if qubit connectivity is a restraint that needs to be accounted for, the procedure above can be altered to account for that by instead disentangling a different set of subsystems~\cite{thesis}. In this case however, an MPS representation may not be as useful in forming the reduced density matrices.   

Lastly, one efficiency that can be made if we are to relax the requirement of exact disentangling is that several disjoint subsystems may be processed at once, this can result in shallower circuits as per Figure~\ref{fig:mps_circ}(b). Note, however that this circuit results in a higher depth circuit over multiple layers.

\section{Quantum Variational Classifiers}

The paradigmatic QML algorithm is the Quantum Variational Classifier (QVC)\footnote{You will sometimes find these referred to as Variational Quantum Classifiers (VQCs) or Quantum Neural Networks (QNNs) in the literature}. The concept behind this algorithm is rather simple, one has some unitary operation $U(\vec{\theta})$ parameterised by some angles $\vec{\theta}$ whereby classification is determined by measuring some observable set of observables $\{O_i\}_{i=0}^m$ where $m$ is the number of classes. Predictably the classification corresponds to the observable with the largest expectation values, i.e.
\begin{equation}
    \hat{y}(v) = \arg \max_j \bra{v}U(\vec{\theta})^\dagger O_j U(\vec{\theta})\ket{v}.
\end{equation}
Typically the parameterised unitary is composed of many repeating layers of local one and two-qubit gates generally chosen such that some notion of locality is maintained in the two-qubit gates, e.g.~\cite{callibrating} or~\cite{hardwareeff}.

Following classical networks, training is performed by assigning a probability of classification, for example using softmax normalisation,
\begin{equation}
    p_i(v) = \frac{\exp{(\bra{v}U(\vec{\theta})^\dagger O_i U(\vec{\theta})\ket{v})}}{\sum_j \exp{(\bra{v}U(\vec{\theta})^\dagger O_j U(\vec{\theta})\ket{v})}}.
\end{equation}
One can then update the parameters $\vec{\theta}$ by minimising a loss function such as the cross-entropy loss,
\begin{equation}
    \mathcal{L} = \frac{1}{N} \sum_{i=1}^N -\log(p_{y(v_i)}(v_i))   
\end{equation}
where we have used the index $i$ to iterate over different inputs $v_i$ and where $y(v_i)$ refers to the true label of $v_i$. The loss function, although determined on a quantum device is minimised using a classical gradient based optimiser such as ADAM~\cite{adam}.

The primary advantage of QVCs is that they are easy to implement and are rather versatile, being able to handle generic classification tasks. However, they are susceptible to barren plateaus~\cite{barren} which significantly hinders the training of deep variational quantum circuits. Luckily though, one may adapt the variational circuit to attempt to mitigate this affect~\cite{rotationally,fixbarren}.

\section{Approximate Encoding in QVCs}

\begin{figure}
    \centering
    \includegraphics[width=0.5\linewidth]{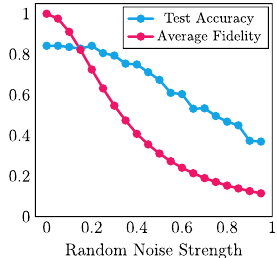}
    \caption{Test Accuracy for 500 images of the MNIST dataset~\cite{mnist} in the presence of random perturbations to the input amplitudes on a trained QVC
    with many layers. Note that the test accuracy only dips after the input fidelity drops below $60\%$. Reprint from~\cite{drastic}}
    \label{fig:rand_peturb}
\end{figure}
\begin{figure}
    \centering
    \includegraphics[width=\linewidth]{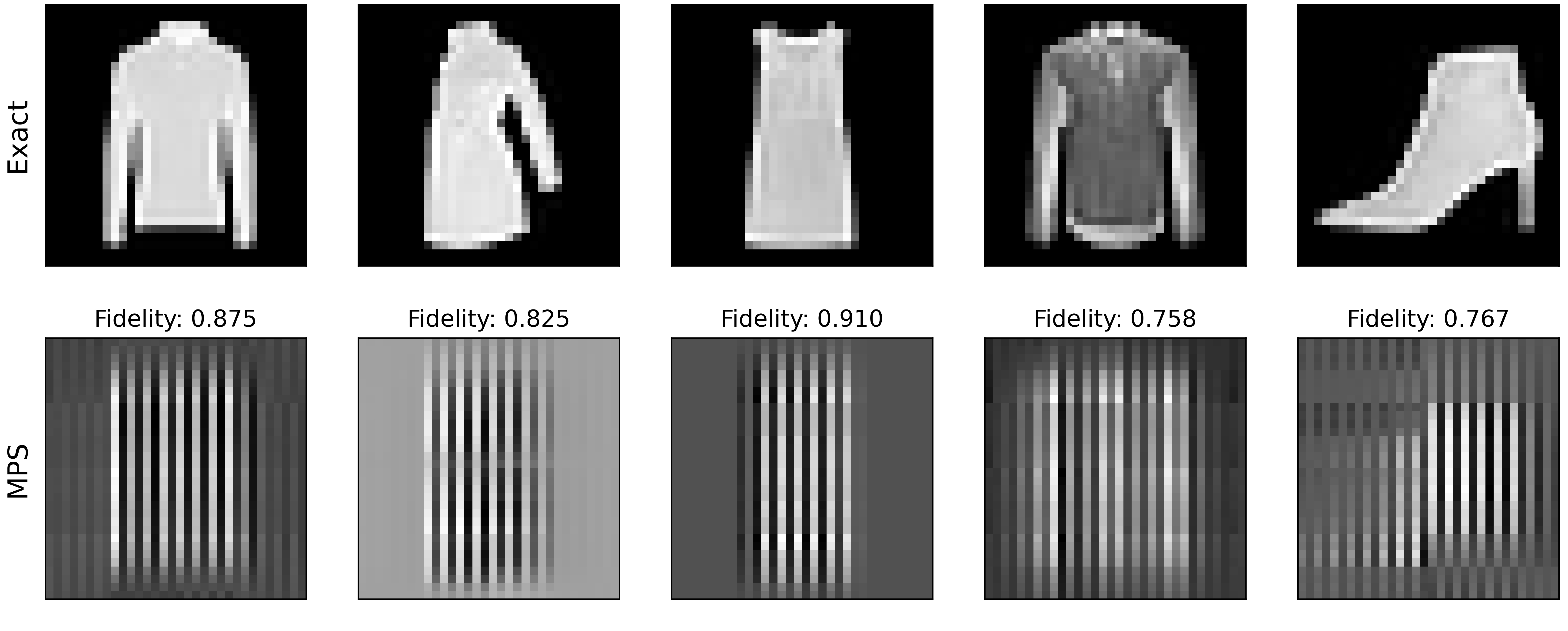}
    \caption{Example images from the Fashion MNIST dataset upon approximate encoding using the MPS assisted protocol. Note that the vertical stripes that result from the encoding as per Equation~\eqref{eq:compr}. Reprint from~\cite{drastic}}
    \label{fig:mnist_mps}
\end{figure}

As it turns out, one does not necessarily need highly accurate input states to perform classification tasks. Intriguingly, as reported in~\cite{drastic} and reproduced in Figure~\ref{fig:rand_peturb}, it appears that classification accuracy appears to be somewhat resilient to random perturbations on the input. On the face of it, this suggests that perhaps an approximate encoding may be sufficient for classification tasks. An important consideration however is that the approximate encoding of the kind that is described above is not akin to random perturbations. For example if we consider the approximate preparation of images in Fashion MNIST dataset using the above procedure as reported in~\cite{drastic}, we see in Figure~\ref{fig:mnist_mps} this results in images that are blocky with alternating vertical stripes. These visual artefacts appear to be a result of the `compressed' encoding of the images,
\begin{equation}
   v_j \mapsto \frac{1}{||v||} \sum_{i=0}^{2^{N-1}-1} (v_{2i} + \text{i}v_{2i+1})\ket{i} \label{eq:compr},
\end{equation}
\begin{figure}
    \centering
    \includegraphics[width=0.5\linewidth]{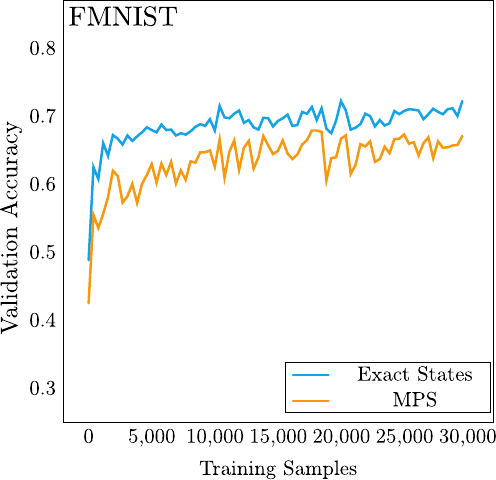}
    \caption{The training accuracy for a deep nine qubit QVC using both an exact encoding and MPS assisted encoding with a threshold accuracy of $60\%$. Figure adapted from Supplemental Material of~\cite{drastic}}
    \label{fig:fmnist_train}
\end{figure}
where one can reduce the number of required qubits by one by encoding adjacent qubits (in this case the horizontal qubits) in the real and imaginary components of the basis state respectively. If one were to use a more conventional amplitude encoding scheme without any compression, one would observe that the approximate encoding is ``blocky'' in nature (see Supplemental Figure 2 of~\cite{drastic}). Nevertheless, we empirically observe that the classification accuracy when using these approximate inputs for both training and testing remains high as these artefacts do appear to be somewhat local. This can be seen in Figure~\ref{fig:fmnist_train} where the MPS assisted encoding method with a threshold fidelity of $60\%$ results in a training accuracy only marginally worse than that achieved when using an exact encoding. 

\begin{figure}
    \centering
    \includegraphics[width=0.8\linewidth]{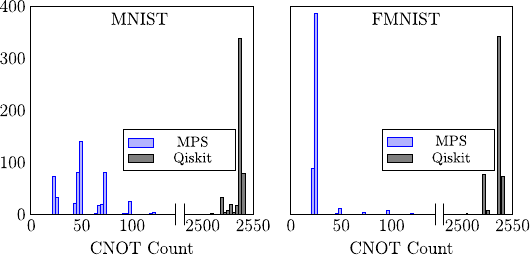}
    \caption{CNOT gate count for $500$ MNIST and FMNIST input states using the MPS method with a target fidelity of $60\%$, as well as the Qiskit~\cite{qiskit} exact state preparation function based off~\cite{shende}. Note that the MPS method results in a circuit with significantly fewer CNOT gates, suggesting that at most only seven iterations of the algorithm are required to rearch the target fidelity. Figure adapted from~\cite{drastic}}
    \label{fig:counts}
\end{figure}

Noting that the MPS assisted encoding runs over a discrete number of iterations and is less heuristic than other approximate encoding schemes, one finds that the true fidelity of the approximately encoded inputs can be rather variable and indeed for the MNIST and FMNIST dataset, it is frequently greater than the target fidelity. This is particularly pronounced for the FMNIST dataset where the fidelity distribution peaks at $83\%$. Consequently the number of gates required, which is a direct function of the number of iterations, is less fixed but is orders of magnitude less than that resulting from conventional state preparation algorithms as per Figure~\ref{fig:counts}.  

\subsection{Robustness against classical adversarial attacks}

One of the interesting developments resulting from the advent of QML is the possibility that QML classifiers are robust towards classically adversaries~\cite{west_benchmarking,west_adversarial}. \textit{A priori} this is not obviously true as adversaries trained on a given classical learner are seemingly transferable to other completely independent learners~\cite{intriguing,exploring}. Intriguingly, there is also evidence towards quantum adversaries being capable of successfully attacking classical networks~\cite{west_benchmarking}. As a result, there may be a niche in which QML may present an advantage wherein security is of utmost importance and quantum deices are not so readily available as to allow an attacker to train a quantum adversary.

The approximate encoding of input states also appears to present an increased robustness towards classical adversarial attacks~\cite{drastic}. In some sense, this is not surprising as random perturbations have been shown to increase the robustness of classical networks~\cite{smooth,noise} as the noisy input classifiers ``wash-out'' the adversarial perturbations that result from training on the exact input. Importantly though, the approximate state preparation discussed here is not stochastic in nature and so in principle one could train an adversary on the approximate inputs. However, given that the approximate input preparation can occur at runtime, adversarial perturbations may again be overcome.

\begin{figure}
    \centering
    \includegraphics[width=0.75\linewidth]{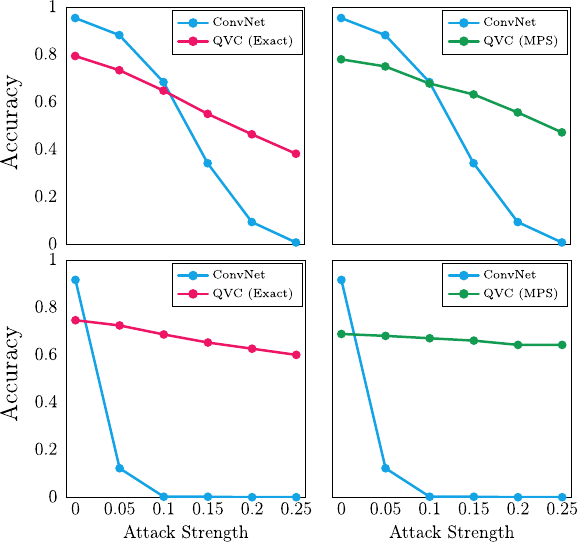}
    \caption{(Top Row) Test accuracies for a trained Convolutional Neural Network (ConvNet) and QVC on adversarially perturbed images from the MNIST dataset. The adversary is trained on the ConvNet classifier with the attack strength corresponding to the step-size in the Projective Gradient Decent (PGD)~\cite{pgd}. Further details on the adversary training can be found in~\cite{drastic}. (Bottom Row) Similar results using images from the FMNIST dataset. For both MNIST and FMNIST it is found that while both the QVC with exact inputs and approximate inputs are robust against the adversarial perturbations, the approximate inputs show an increased robustness towards the perturbations. Figure adapted from~\cite{drastic}.}
    \label{fig:adv}
\end{figure}

The adversarial examples reported in~\cite{drastic} (and reproduced in Figure~\ref{fig:adv}) for the MNIST and FMNIST dataset show that QVCs trained and tested on approximate inputs show an increased test accuracy compared to QVCs with exact inputs. This improvement, although notable is in stark contrast to the accuracy of the classical convolutional neural network which drops off considerably under adversarial attack.

\subsection{Experimental demonstration on a superconducting quantum computer}

Although quantum devices are incapable of running moderately deep circuits with high fidelity, one can experimentally demonstrate successful classification on a quantum device by considering simple datasets on few qubits. For example the toy ``Shapes'' dataset considered in~\cite{drastic} consists of $8\times 8$ images, requiring five qubits to encode, and only has two classes. This dataset was evaluated on the \texttt{ibm\_algeirs} device with the Qiskit~\cite{qiskit} exact state preparation method adapted from~\cite{shende} achieving an average test accuracy of $45.8\%\pm6.7$ compared to the MPS assisted method which achieved $95\%\pm2.3$. One also continues to see a robustness towards adversarial perturbations, albeit with a diminished test accuracy of $66.8\% \pm 0.3$. More results are presented in Figure 3 of~\cite{drastic}.

\section{Conclusion}

As more sophisticated QML algorithm continue to be developed, the input data and its encoding will remain a constant presence. In the context of specific classification tasks one can always develop a bespoke encoding that relies on the underlying structure and symmetry of the task at hand~\cite{provably,jamie}. For general classification tasks however, one will not have the ability to rely on this and in that context it remains crucial to be able to encode data efficiently in such a way that does not negate the possible exponential advantages that arise from QML algorithms. To that end, one may be able to utilise the inherent noise resilience of QML algorithms to encode states in a lossy fashion. This comes with the added advantage that such encodings lend themselves to being resilient towards classical adversarial attacks that are commonly trained using an exact representation of the input. An open question remains whether such an advantage exists when adversaries also have access to quantum devices or indeed have prior knowledge as to the encoding algorithm.  

\section*{Acknowledgements}
We thank our collaborators Jamie Heredge, Floyd Creevey, Martin Sevior and Lloyd Hollenberg with whom the results presented this chapter were originally published in~\cite{drastic}. C.N. and M.W. acknowledge the
support of Australian Government Research Training Program
Scholarship 

%
%
%

\end{document}